\documentclass[12pt]{iopart}
\usepackage{graphicx}

\begin{document}

\title[Teleportation of an arbitrary multipartite state...]{Teleportation of
an arbitrary multipartite state via photonic Faraday rotation}
\author{Juan-Juan Chen$^{1,2}$, Jun-Hong An$^{1}$\footnote{%
Email: anjhong@lzu.edu.cn}, Mang Feng$^{3}$ and Ge Liu$^{2}$}
\address{$^{1}$ Center for Interdisciplinary Studies, Lanzhou University, Lanzhou 730000,
China}
\address{$^{2}$ Department of Modern Physics, Lanzhou
University, Lanzhou 730000, China}
\address{$^{3}$ State Key Laboratory of Magnetic Resonance and Atomic and
Molecular Physics, Wuhan Institute of Physics and Mathematics,
Chinese Academy of Sciences, Wuhan 430071, China}

\begin{abstract}
We propose a practical scheme for deterministically teleporting an arbitrary
multipartite state, either product or entangled, using Faraday rotation of
the photonic polarization. Our scheme, based on the input-output process of
single-photon pulses regarding cavities, works in low-Q cavities and only
involves virtual excitation of the atoms, which is insensitive to both
cavity decay and atomic spontaneous emission. Besides, the Bell-state
measurement is accomplished by the Faraday rotation plus product-state
measurements, which could much relax the experimental difficulty to realize
the Bell-state measurement by the CNOT operation.
\end{abstract}

\maketitle



\section{Introduction}

\label{Intod} Teleportation is the faithful transfer of quantum states
between spatially separated parties, based on the prior establishment of
entanglement and a classical communication \cite{Bennett93}. As a practical
application, teleportation has been implemented experimentally in several
quantum systems \cite{Bouwmeester97,Sherson06}, and could be very useful for
achieving quantum repeaters \cite{Briegel98}, quantum networks \cite%
{Kimble08}, and also quantum computing \cite{Gottesman99}. Recently, much
attention has been paid on teleportation of multipartite states \cite%
{Lee02,Fang03,Rigolin05Yeo06Muralidharan08}. It has been shown that an
arbitrary two-qubit state can be teleported using the multipartite entangled
state as quantum channel \cite{Rigolin05Yeo06Muralidharan08}. On the other
hand, cavity QED system gives a very nice platform to accomplish quantum
teleportation. Many schemes have been proposed in this system. The earlier
schemes use the atom as a flying qubit to transfer quantum information \cite%
{Liu04,Zheng04Cardoso05}, which are actually unsuitable for long-distance
communication. Furthermore, the schemes mentioned above work well only in
high-Q cavities, which are hard to accomplish with current technology.
Besides, most of those schemes, e.g., \cite{Fang03,Zheng04Cardoso05}, are
intrinsically probabilistic.

Can we achieve teleportation with low-Q cavities? More recently, schemes
using photons to transfer information and atoms to store information were
proposed \cite{Bose99}. These schemes are based on the detection of photons
leaking out of cavities. Taking the cavity damping into account, these
schemes work in the low-Q cavity regime. However, the cavity damping
actually plays a detrimental role in the success probability of the
teleportaion. Thus these schemes are intrinsically probabilistic. It shows
the strong dependence of the implementation on the dissipative rate. This
implies a negligible possibility to achieve a teleportation with the low-Q
cavities.

In this work, however, we will show this possibility with a practical scheme
to teleport multipartite states using sophisticated low-Q cavities. Inspired
by recent experiments of photonic input-output process \cite%
{Sherson06,Dayan08}, the key idea is to make use of the Faraday rotation
produced by single-photon-pulse input and output process regarding low-Q
cavities. We had noticed a previous idea for teleportation using Faraday
rotation by single photons from microcavities at one location to those at
another location in quantum dot system \cite{Leuenberger05}. While the
photons input and output from the microcavities correspond to some decay
effects, there was little discussion about the impact from dissipation on
the teleportation. In our point of view, an efficient implementation of the
teleportation in \cite{Leuenberger05} requires high rate of the photons
input and output with respect to the cavities. This implies the necessity of
employing bad cavities with large decay rates, which goes beyond the model
there.

The key point of our present scheme is to make use of a new reflection
coefficient equation published recently \cite{An09}, which enables us to get
a rotation of the photonic polarization, named Faraday rotation \cite%
{Julsgaard01}, conditional on the atomic state confined in the cavity even
in low-Q regime. We argue that our scheme would be advantageous over
previous schemes \cite{Bose99,Leuenberger05} in accomplishment of the
teleportation in a deterministic fashion with currently available cavity QED
technology, such as the microtoroidal resonator \cite{Dayan08} or the
single-sided optical cavity \cite{Turchette95}. Furthermore, due to the
disentangling action of Faraday rotation to the photonic and atomic state in
the destination side, only product-state measurement is needed in our
scheme, which much relaxes the experimental difficulty.

This paper is organized as follows. In Sec. \ref{fd}, we first briefly
review the input-output process of a single-photon pulse regarding a low-Q
cavity and show the Faraday Rotation of the photonic polarization \cite{An09}%
. In Sec. \ref{sche} we explicitly elucidate our teleportation scheme of
multipartite state via Faraday rotation. The discussion about the
experimental feasibility of our scheme and a short conclusion are made in
Sec. \ref{dc}.

\begin{figure}[t]
\begin{center}
\scalebox{0.5}{\includegraphics{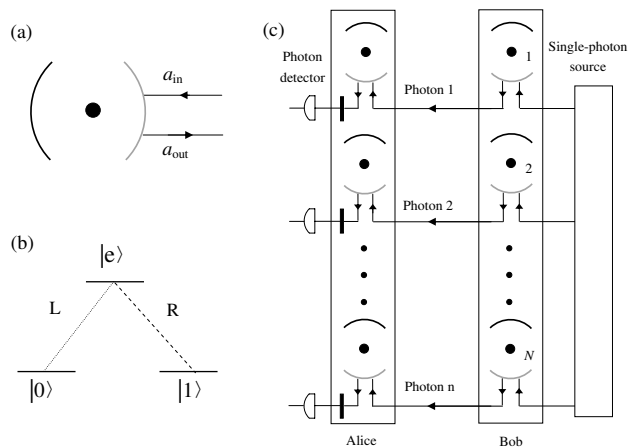}}
\end{center}
\caption{(a) Single photon input and output regarding the low-Q cavity. (b)
The level structure of the atom confined in the cavity. The transitions from
the degenerate doublet ground states $|0\rangle$ and $|1\rangle$ to the
excited state $|e\rangle$ are triggered by a L - and R - circularly
polarized photons, respectively. (c) Schematic for teleporting an unknown
multipartite state of atoms, where the arrows show the flying direction of
the photons, and the bold lines are QWPs defined in the text.}
\label{sle}
\end{figure}

\section{Faraday rotation in cavity QED system}

\label{fd} We first briefly review the input-output relation in a low-Q
cavity discussed in \cite{An09}. Consider a bimodal cavity with one of the
mirrors partially reflective so that the photon can be injected in and then
reflected out with a certain probability, as shown in Fig. \ref{sle} (a).
The level structure of the atom confined in the cavity is depicted in Fig. %
\ref{sle} (b), where $|0\rangle$ and $|1\rangle$ correspond to the
degenerate Zeeman sublevels of the ground state in an alkali atom, and $%
|e\rangle$ is the excited state. The atomic transitions $|e\rangle%
\leftrightarrow|0\rangle$ and $|e\rangle\leftrightarrow|1\rangle$ are due to
the coupling of the atom to the two degenerate cavity modes $a_L$ (with left
circular polarization) and $a_R$ (with right circular polarization),
respectively. Each of the transition channels is governed by the Hamiltonian
of usual Jaynes-Cummings model.

Suppose a L - (or R -) circularly polarized single-photon pulse with the
frequency $\omega_p$ is input into the cavity. The pulse is denoted by $%
|\psi\rangle=\int_0^Tf(t)a^\dag_{in}(t)dt|vac\rangle$, where $f(t)$ is a
time-dependent normalized pulse shape with the pulse duration $T$, $%
a^\dag_{in}$ is the input photon operator satisfying the commutation
relation $[a_{in}(t),a^\dag_{in}(t^{\prime })]=\delta(t-t^{\prime })$, and $%
|vac\rangle$ is the vacuum state of the optical field \cite{Duan04}. If the
atom is prepared in $|0\rangle$ (or in $|1\rangle $) initially, then the
photon will trigger the transition $|0\rangle\leftrightarrow|e\rangle$ (or $%
|1\rangle\leftrightarrow|e\rangle$). The quantum Langevin equations are
\begin{eqnarray}
\dot{a}(t)&=&-[i(\omega _{c}-\omega _{\mathrm{p}})+\frac{\kappa }{2}%
]a(t)-g\sigma _{-}(t)-\sqrt{\kappa }a_{\mathrm{in}}(t),  \nonumber \\
\dot{\sigma}_{-}(t)&=&-[i(\omega _{0}-\omega _{\mathrm{p}})+\frac{\gamma }{2}%
]\sigma _{-}(t)-g\sigma _{z}(t)a(t) +\sqrt{\gamma }\sigma _{z}(t)b_{\mathrm{%
in}}(t),  \label{eom2}
\end{eqnarray}%
where $\kappa $ and $\gamma$ are the cavity damping rate and atomic
spontaneous emission rate, respectively. $b_{in}(t)$ is vacuum input field
felt by the atom. The contribution of vacuum field $b_{in}(t)$ is negligible
because it is much less than the one of the photon pulse $a_{in}(t)$. The
output field $a_{out}(t)$ relates to the input field by the intracavity
field as $a_{out}(t)=a_{in}(t)+\sqrt{\kappa }a(t)$ \cite{Walls94}. Assume
that $\kappa$ is large enough to ensure the atom initially prepared in the
ground state is only virtually excited by the photon, so we can take $%
\langle \sigma_z(t)\rangle=-1$. Also under this large $\kappa$ limit, we can
adiabatically eliminate the intracavity mode from the set of quantum
Langevin equations and arrive consequently at the input-output relation as
\begin{equation}
r(\omega_p)=\frac{[i(\omega _{c}-\omega _{p})-\frac{\kappa }{2}][i(\omega
_{0}-\omega _{p})+\frac{\gamma }{2}]+g^{2}}{[i(\omega _{c}-\omega _{p})+%
\frac{\kappa }{2}][i(\omega _{0}-\omega _{p})+\frac{\gamma }{2}]+g^{2}}.
\label{r}
\end{equation}%
where $r(\omega _{p})=\frac{a_{out}(t)}{a_{in}(t)}$ is the reflection
coefficient of the photon to the atom-cavity system. On the other hand, if
the atom is prepared in $|1\rangle$ (or $|0\rangle$) initially under the
condition that a L- (or R-) circularly polarized photon pulse is input in
the cavity, then no transition would be triggered. In other words, the
photon only feels an empty cavity, for which the input-output relation
corresponds to Eq. (\ref{r}) with $g=0$ \cite{Walls94}, i.e.
\begin{equation}
r_0(\omega_p)=\frac{i(\omega _{c}-\omega _{p})-\frac{\kappa }{2}}{i(\omega
_{c}-\omega _{p})+\frac{\kappa }{2}}.  \label{r0}
\end{equation}

The complex reflection coefficients (\ref{r}) and (\ref{r0}) show that the
reflected photon experiences an absorption as well as a phase shift denoted
by $e^{i\phi}$ and $e^{i\phi_0}$, respectively. However, under the practical
experimental condition, i.e., the strong $\kappa$ and weak $g$ and $\gamma$
\cite{Dayan08}, it has been verified that the absolute values of $r(\omega_p)
$ and $r_0(\omega_p)$ are only slightly deviated from unity \cite{Hu08,An09}%
. This implies that the photon experiences a very weak absorption, and
thereby we may approximately consider that the output photon only
experiences a pure phase shift without any absorption.

With this basic input-output relation, the Faraday rotation can be derived
readily when a linearly polarized photon is input into the cavity. Consider
the input pulse is linearly polarized in $|\Psi_{in}\rangle=\frac{1}{ \sqrt{2%
}}(|L\rangle+|R\rangle)$ and the atom initially in $|0\rangle$, the output
photon would have a rotation in the polarization. In this case, the $%
|L\rangle$ component of the photon virtually triggers the transition of the
atom from $|0\rangle$ to $|e\rangle$, and thus experiences a phase shift $%
e^{i\phi}$ obeying Eq. (\ref{r}). In contrast, the $|R\rangle$ component of
the photon only feels an empty cavity, which yields a phase shift $%
e^{i\phi_0}$ obeying Eq. (\ref{r0}). So the output pulse is
\begin{equation}
|\Psi_{out}\rangle_-=\frac{1}{\sqrt{2}}(e^{i\phi}|L\rangle+e^{i\phi_0}|R
\rangle).  \label{po1}
\end{equation}
This also implies that the polarization direction of the reflected photon
rotates an angle $\Theta^-_F=\frac{\phi_0-\phi}{2}$ with respect to the
input one, called Faraday rotation \cite{Julsgaard01}. Similarly, if the
atom is initially prepared in $|1\rangle$, then only the R circularly
polarized photon could sense the atom, while the L circularly polarized
photon only feels the empty cavity. So we have
\begin{equation}
|\Psi_{out}\rangle_+=\frac{1}{\sqrt{2}}(e^{i\phi_0}|L\rangle+e^{i\phi}|R
\rangle),  \label{po2}
\end{equation}
corresponding to a Faraday rotation with an angle $\Theta_F^+=\frac{
\phi-\phi_0}{2}$.

\section{The realization of multipartite-state teleportation via Faraday
rotation}

\label{sche} In what follows, we show the teleportation of a
multipartite state by above Faraday rotations. As plotted in Fig. 1
(c), to construct the entanglement between Alice and Bob, Bob has to
first input $N$ single-photon pulses into his cavities to produce
$N$ pairs of entangled states between the photons and atoms using
the Faraday rotations of the photons. Then the emitted photons fly
to Alice through the fiber. Once Alice collects the photons in her
cavities, the entanglement between Alice and Bob has been
established. We emphasize that the photon in our scheme acts dual
roles: On the one hand, it acts as a bus to distribute entanglement;
On the other hand, it also acts as a component of the entangled pair
to implement the teleportation.

\subsection{The case of bipartite state}
To demonstrate our scheme specifically, we will take $N=2$ below as
an example. Suppose the quantum state to be teleported in Alice's
hands is \cite {explain0} \begin{equation} \left\vert \varphi
\right\rangle =\alpha \left\vert 01\right\rangle +\beta \left\vert
10\right\rangle +\zeta \left\vert 00\right\rangle +\delta \left\vert
11\right\rangle ,
\end{equation}
where $\alpha $, $\beta $, $\zeta $, and $\delta $ are unknown parameters
with $\alpha ^{2}+\beta ^{2}+\zeta ^{2}+\delta ^{2}=1$. In the following we
give explicitly the steps for accomplishing the quantum teleportation.

\emph{The first step: Establishment of the quantum channels.} Bob
inputs two linearly polarized single-photon pulses in $\left\vert
\Psi \right\rangle _{i}=\frac{1}{\sqrt{2}}\left( \left\vert
L\right\rangle _{i}+\left\vert R\right\rangle _{i}\right) $
generated from the single-photon source into two cavities at his
side. The states of the atoms confined in Bob's cavities are
initially $|\psi \rangle _{i}=\frac{1}{\sqrt{2}}(|0\rangle
_{i}+|1\rangle _{i}),~(i=1,2)$. According to the input-output
relation, the atoms and the photons are entangled due to the Faraday
rotation,
\begin{equation}
\left\vert \Psi \right\rangle _{i}\left\vert \psi \right\rangle
_{i}\rightarrow \frac{1}{\sqrt{2}}[|0\rangle _{i}|\Psi _{out}\rangle
_{-i}+|1\rangle _{i}|\Psi _{out}\rangle _{+i}].  \label{qch}
\end{equation}
Then after Alice collects the two output photons via two fibers, two
quantum channels are thus established. One can see that $|\Psi
_{out}\rangle _{+}$ and $|\Psi _{out}\rangle _{-}$ are orthogonal
when $\phi-\phi_0=\pi/2$. It means that under this condition, the
quantum channel characterized by the right side of Eq. (\ref{qch})
is maximally entangled. In practice, this condition can be easily
satisfied experimentally. Using the parameters in Ref.
\cite{Dayan08}, $\omega_0=\omega_c$,
$\omega_p=\omega_c-\frac{\kappa}{2}$, and $g=\frac{\kappa}{2}$, we
can verify from Eqs. (\ref{r}) and (\ref{r0}) that $\phi=\pi$ and
$\phi_0=\pi/2$. The Faraday rotation under the same condition has
been used to realize the universal quantum gate, the entanglement
generation between remote atoms and its conversion to the flying
photons in Ref. \cite{An09,Chen09}.

\emph{The second step: Realization of the Bell-state measurement.}
In the standard protocol of quantum teleportaion, a necessary step
is the Bell-state measurement on the atom and one of the entangled
pair of the quantum channel in Alice's side \cite{Bennett93}. The
measurement would collapse the state of the total system in one of
its four superposition components. Experimentally, this Bell-state
measurement is ordinarily realized by the CNOT operation plus the
product-state measurement. In the following we show the CNOT
operation can actually be replaced by the Faraday rotation in
Alice's side, which much relaxes the experimental difficulty to do
the CNOT operation. After collecting the photons, Alice feeds the
photons to her cavities. The Faraday rotation makes the two atoms at
Alice's hands entangled with the two quantum channels. So the state
of the entire system could be written as,
\begin{eqnarray}
&&|\varphi \rangle \prod_{i=1,2}\frac{1}{\sqrt{2}}[|0\rangle _{i}|\Psi
_{out}\rangle _{-i}+|1\rangle _{i}|\Psi _{out}\rangle _{+i}]\rightarrow
\nonumber \\
&&\frac{1}{4}[\left\vert LL\right\rangle \left( -i\alpha \left\vert
01\right\rangle -i\beta \left\vert 10\right\rangle +\zeta \left\vert
00\right\rangle -\delta \left\vert 11\right\rangle \right)   \nonumber \\
&&\times \left( \left\vert 00\right\rangle _{12}-i\left\vert 01\right\rangle
_{12}-i\left\vert 10\right\rangle _{12}-\left\vert 11\right\rangle
_{12}\right)   \nonumber \\
&&+\left\vert LR\right\rangle \left( \alpha \left\vert 01\right\rangle
-\beta \left\vert 10\right\rangle -i\zeta \left\vert 00\right\rangle
-i\delta \left\vert 11\right\rangle \right)   \nonumber \\
&&\times \left( -i\left\vert 00\right\rangle _{12}+\left\vert
01\right\rangle _{12}-\left\vert 10\right\rangle _{12}-i\left\vert
11\right\rangle _{12}\right)   \nonumber \\
&&+\left\vert RL\right\rangle \left( -\alpha \left\vert 01\right\rangle
+\beta \left\vert 10\right\rangle -i\zeta \left\vert 00\right\rangle
-i\delta \left\vert 11\right\rangle \right)   \nonumber \\
&&\times \left( -i\left\vert 00\right\rangle _{12}-\left\vert
01\right\rangle _{12}+\left\vert 10\right\rangle _{12}-i\left\vert
11\right\rangle _{12}\right)   \nonumber \\
&&+\left\vert RR\right\rangle \left( -i\alpha \left\vert 01\right\rangle
-i\beta \left\vert 10\right\rangle -\zeta \left\vert 00\right\rangle +\delta
\left\vert 11\right\rangle \right)   \nonumber \\
&&\times \left( -\left\vert 00\right\rangle _{12}-i\left\vert
01\right\rangle _{12}-i\left\vert 10\right\rangle _{12}+\left\vert
11\right\rangle _{12}\right) ],  \label{secst}
\end{eqnarray}%
where the kets with and without subscripts correspond to the atomic
states at Bob's and Alice's hands, respectively. We have used $\phi
=\pi $ and $\phi _{0}=\pi /2$ with the practical parameters in Ref.
\cite{Dayan08}.

Next, Alice makes Hadamard gates on the photons and atoms at her
side. The photonic Hadamard gating is realized by a quarter-wave
plate (QWP), which makes $|R\rangle\rightarrow
\frac{|L\rangle-|R\rangle}{\sqrt{2}}$ and $|L\rangle\rightarrow
\frac{|L\rangle+|R\rangle}{\sqrt{2}}$. Then the right-hand side of
Eq. (\ref{secst}) is converted into $\sum_{i,j;m,n}|ij, mn \rangle
|f_{ij,mn}\rangle_{12}$, where $|ij\rangle$, $|mn\rangle$, and
$|f_{ij,mn}\rangle_{12}$ denote the photon, and Alice's and Bob's
atomic states, respectively, as shown in Table \ref{tab}.

Finally, Alice performs measurement on the states of the photons and the
atoms at her side, followed by the collapse of the atomic state at Bob's
side to one of the corresponding components in above superposition.

\emph{The third step: Bob's recovery operations.} Based on the
message $(i,j,m,n)$ from Alice via the classical channel about her
measurement, Bob could deterministically recover the unknown state
only by some local operations $M_{ij,mn}$ on his atoms, as listed in
Table \ref{tab}.

\begin{table}[tbp]
\caption{The superposition components of the final state and Bob's
corresponding operations when $N=2$.} \label{tab}\tabcolsep 5mm
\par
\begin{center}
\begin{tabular}{ccc}
\hline\hline $|ij,mn\rangle$ & $|f_{ij,mn}\rangle_{12}$ &
$M_{ij,mn}$ \\ \hline $\left\vert LL\right\rangle \left\vert
00\right\rangle $ & $-\alpha \left\vert 10\right\rangle -\beta
\left\vert 01\right\rangle -\zeta
\left\vert 11\right\rangle -\delta \left\vert 00\right\rangle$ & $%
-\sigma_x^{(1)}\otimes\sigma_x^{(2)}$ \\
$\left\vert LL\right\rangle \left\vert 11\right\rangle $ & $\alpha
\left\vert 10\right\rangle+\beta \left\vert 01\right\rangle-\zeta
\left\vert 11\right\rangle-\delta \left\vert 00\right\rangle$ & $%
\sigma_y^{(1)}\otimes\sigma_y^{(2)}$ \\
$\left\vert LL\right\rangle \left\vert 01\right\rangle $ & $-\alpha
\left\vert 10\right\rangle+\beta \left\vert 01\right\rangle +\zeta
\left\vert11\right\rangle -\delta \left\vert 00\right\rangle $ & $%
\sigma_x^{(1)}\otimes\sigma_y^{(2)}$ \\
$\left\vert LL\right\rangle \left\vert 10\right\rangle $ & $\alpha
\left\vert 10\right\rangle -\beta \left\vert 01\right\rangle +\zeta
\left\vert 11\right\rangle -\delta \left\vert 00\right\rangle$ & $%
\sigma_y^{(1)}\otimes\sigma_x^{(2)}$ \\
$\left\vert LR\right\rangle \left\vert 00\right\rangle $ & $i\alpha
\left\vert 11\right\rangle -i\beta \left\vert 00\right\rangle
-i\zeta
\left\vert 10\right\rangle +i\delta \left\vert 01\right\rangle $ & $%
\sigma_x^{(1)}\otimes\sigma_z^{(2)}$ \\
$\left\vert LR\right\rangle \left\vert 11\right\rangle $ & $-i\alpha
\left\vert 11\right\rangle +i\beta \left\vert 00\right\rangle
-i\zeta
\left\vert 10\right\rangle +i\delta \left\vert 01\right\rangle$ & $%
\sigma_y^{(1)}\otimes I^{(2)}$ \\
$\left\vert LR\right\rangle \left\vert 01\right\rangle $ & $i\alpha
\left\vert 11\right\rangle +i\beta \left\vert 00\right\rangle
+i\zeta
\left\vert 10\right\rangle +i\delta \left\vert 01\right\rangle$ & $%
\sigma_x^{(1)}\otimes I^{(2)}$ \\
$\left\vert LR\right\rangle \left\vert 10\right\rangle $ & $-i\alpha
\left\vert 11\right\rangle -i\beta \left\vert 00\right\rangle
+i\zeta
\left\vert 10\right\rangle +i\delta \left\vert 01\right\rangle$ & $%
\sigma_y^{(1)}\otimes\sigma_z^{(2)}$ \\
$\left\vert RL\right\rangle \left\vert 00\right\rangle $ & $i\alpha
\left\vert 00\right\rangle +i\beta \left\vert 11\right\rangle
-i\zeta
\left\vert 01\right\rangle +i\delta \left\vert 10\right\rangle$ & $%
\sigma_z^{(1)}\otimes\sigma_x^{(2)}$ \\
$\left\vert RL\right\rangle \left\vert 11\right\rangle $ & $i\alpha
\left\vert 00\right\rangle-i\beta \left\vert 11\right\rangle -i\zeta
\left\vert 01\right\rangle +i\delta \left\vert 10\right\rangle$ & $%
I^{(1)}\otimes\sigma_y^{(2)}$ \\
$\left\vert RL\right\rangle \left\vert 01\right\rangle $ & $-i\alpha
\left\vert 00\right\rangle-i\beta \left\vert 11\right\rangle +i\zeta
\left\vert 01\right\rangle +i\delta \left\vert 10\right\rangle$ & $%
\sigma_z^{(1)}\otimes\sigma_y^{(2)}$ \\
$\left\vert RL\right\rangle \left\vert 10\right\rangle $ & $i\alpha
\left\vert 00\right\rangle +i\beta \left\vert 11\right\rangle
+i\zeta
\left\vert 01\right\rangle +i\delta \left\vert 10\right\rangle$ & $%
I^{(1)}\otimes\sigma_x^{(2)}$ \\
$\left\vert RR\right\rangle \left\vert 00\right\rangle $ & $-\alpha
\left\vert 00\right\rangle -\beta \left\vert 11\right\rangle +\zeta
\left\vert 01\right\rangle +\delta \left\vert 10\right\rangle $ & $%
\sigma_z^{(1)}\otimes\sigma_z^{(2)}$ \\
$\left\vert RR\right\rangle \left\vert 11\right\rangle $ & $\alpha
\left\vert 00\right\rangle +\beta \left\vert 11\right\rangle +\zeta
\left\vert 01\right\rangle +\delta \left\vert 10\right\rangle$ & $%
I^{(1)}\otimes I^{(2)}$ \\
$\left\vert RR\right\rangle \left\vert 01\right\rangle $ & $-\alpha
\left\vert 00\right\rangle +\beta \left\vert 11\right\rangle -\zeta
\left\vert 01\right\rangle +\delta \left\vert 10\right\rangle$ & $%
\sigma_z^{(1)}\otimes I^{(2)}$ \\
$\left\vert RR\right\rangle \left\vert 10\right\rangle $ & $\alpha
\left\vert 00\right\rangle -\beta \left\vert 11\right\rangle -\zeta
\left\vert 01\right\rangle+\delta \left\vert 10\right\rangle$ & $
I^{(1)}\otimes\sigma_z^{(2)}$ \\ \hline\hline
\end{tabular}
\end{center}
\end{table}

\subsection{The case of tripartite state}
With more single-photon pulses, we may generalize our scheme
straightforwardly to teleportation of bigger entangled states. For
simplicity, we take a tripartite state as an example. Suppose the
teleported tripartite state to be of the GHZ-like form,
\begin{equation}
|\varphi\rangle=\alpha|000\rangle+\delta|111\rangle.
\end{equation}
In the first step, Bob inputs three photons in his cavities. The
input-output process in his side makes the photons entangled with
the atoms in each of the cavities. After Alice collects the three
photons, three quantum channels have thus been established. In the
second step, the Faraday rotation in Alice's side makes the state of
the whole system transform into
\begin{eqnarray}
&&|\varphi \rangle \prod_{i=1,2,3}\frac{1}{\sqrt{2}}[|0\rangle
_{i}|\Psi _{out}\rangle _{-i}+|1\rangle _{i}|\Psi _{out}\rangle
_{+i}]\rightarrow
\nonumber \\
&&\frac{1}{8}[(-|0\rangle _{1}+i|1\rangle _{1})(-|0\rangle
_{2}+i|1\rangle _{2})(-|0\rangle _{3}+i|1\rangle
_{3})|LLL\rangle _{123}(-\alpha |000\rangle -i\delta |111\rangle )  \nonumber \\
&&+(-|0\rangle _{1}+i|1\rangle _{1})(-|0\rangle _{2}+i|1\rangle
_{2})(i|0\rangle _{3}-|1\rangle _{3})|LLR\rangle _{123}(i\alpha
|000\rangle
+\delta |111\rangle ) \nonumber\\
&&+(-|0\rangle _{1}+i|1\rangle _{1})(i|0\rangle _{2}-|1\rangle
_{2})(-|0\rangle _{3}+i|1\rangle _{3})|LRL\rangle _{123}(i\alpha
|000\rangle
+\delta |111\rangle ) \nonumber\\
&&+(-|0\rangle _{1}+i|1\rangle _{1})(i|0\rangle _{2}-|1\rangle
_{2})(i|0\rangle _{3}-|1\rangle _{3})|LRR\rangle _{123}(\alpha
|000\rangle
+i\delta |111\rangle ) \nonumber\\
&&+(i|0\rangle _{1}-|1\rangle _{1})(-|0\rangle _{2}+i|1\rangle
_{2})(-|0\rangle _{3}+i|1\rangle _{3})|RLL\rangle _{123}(i\alpha
|000\rangle
+\delta |111\rangle ) \nonumber\\
&&+(i|0\rangle _{1}-|1\rangle _{1})(-|0\rangle _{2}+i|1\rangle
_{2})(i|0\rangle _{3}-|1\rangle _{3})|RLR\rangle _{123}(\alpha
|000\rangle
+i\delta |111\rangle ) \nonumber\\
&&+(i|0\rangle _{1}-|1\rangle _{1})(i|0\rangle _{2}-|1\rangle
_{2})(-|0\rangle _{3}+i|1\rangle _{3})|RRL\rangle _{123}(\alpha
|000\rangle +i\delta |111\rangle ),  \label{trip}
\end{eqnarray}which could be further transformed into the final form
$\sum_{ijk,lmn}|ijk,lmn\rangle|f_{ijk,lmn}\rangle_{123}$ by local
Hadamard operations applied to the photonic and to the atomic states
in Alice's side. The explicit form of the superposition components
in the final state is listed in Table \ref{tab2}. Finally, Alice
performs measurement on the state of photons and the atoms and then
tells Bob her results. In the last step, Bob has to choose local
operations $M_{ijk,lmn}$ appropriately, as shown in Table
\ref{tab2}, to recover the unknown teleported state on his atoms.

\begin{table}[tbp]
\caption{The superposition components of the final state and Bob's
corresponding operations when $N=3$.} \label{tab2}\tabcolsep 2mm
\par
\begin{center}
\begin{tabular}{cccc}
\hline\hline $|ijk\rangle$ & $|lmn\rangle$ & $|f_{ijk,lmn}\rangle_{123}$ & $M_{ijk,lmn}$ \\
\hline $|LLL\rangle$ &
\begin{tabular}{l}
$|000\rangle,|011\rangle,|101\rangle,|110\rangle$ \\
$|001\rangle,|010\rangle,|100\rangle,|111\rangle$%
\end{tabular}
&
\begin{tabular}{l}
$i\alpha |111\rangle +i\delta |000\rangle $ \\
$i\alpha |111\rangle -i\delta |000\rangle $%
\end{tabular}
& \begin{tabular}{l}
$\sigma_x\otimes\sigma_x\otimes\sigma_x$ \\
$\sigma_x\otimes\sigma_x\otimes\sigma_y$%
\end{tabular} \\
$|RRR\rangle$ &
\begin{tabular}{l}
$|000\rangle,|011\rangle,|101\rangle,|110\rangle$ \\
$|001\rangle,|010\rangle,|100\rangle,|111\rangle$%
\end{tabular}
&
\begin{tabular}{l}
$\alpha |000\rangle -\delta |111\rangle $ \\
$\alpha |000\rangle +\delta |111\rangle $%
\end{tabular}
& \begin{tabular}{l}
$I\otimes I\otimes\sigma_z$ \\
$I\otimes I\otimes I$%
\end{tabular} \\
$|LLR\rangle$ &
\begin{tabular}{l}
$|000\rangle,|011\rangle,|101\rangle,|110\rangle$ \\
$|001\rangle,|010\rangle,|100\rangle,|111\rangle$%
\end{tabular}
&
\begin{tabular}{l}
$-\alpha |110\rangle +\delta |001\rangle $ \\
$-\alpha |110\rangle -\delta |001\rangle $%
\end{tabular}
& \begin{tabular}{l}
$\sigma_x\otimes\sigma_x\otimes\sigma_z$ \\
$\sigma_x\otimes\sigma_x\otimes I$%
\end{tabular} \\
$|LRL\rangle$ &
\begin{tabular}{l}
$|000\rangle,|011\rangle,|101\rangle,|110\rangle$ \\
$|001\rangle,|010\rangle,|100\rangle,|111\rangle$%
\end{tabular}
&
\begin{tabular}{l}
$-\alpha |101\rangle +\delta |010\rangle $ \\
$-\alpha |101\rangle -\delta |010\rangle $%
\end{tabular}
& \begin{tabular}{l}
$\sigma_x\otimes\sigma_z\otimes\sigma_x$ \\
$\sigma_x\otimes I \otimes\sigma_y$%
\end{tabular} \\
$|RLL\rangle$ &
\begin{tabular}{l}
$|000\rangle,|011\rangle,|101\rangle,|110\rangle$ \\
$|001\rangle,|010\rangle,|100\rangle,|111\rangle$%
\end{tabular}
&
\begin{tabular}{l}
$-\alpha |011\rangle +\delta |100\rangle $ \\
$-\alpha |011\rangle -\delta |100\rangle $%
\end{tabular}
& \begin{tabular}{l}
$\sigma_z\otimes\sigma_x\otimes\sigma_x$ \\
$I\otimes\sigma_x\otimes\sigma_y$%
\end{tabular} \\
$|LRR\rangle$ &
\begin{tabular}{l}
$|000\rangle,|011\rangle,|101\rangle,|110\rangle$ \\
$|001\rangle,|010\rangle,|100\rangle,|111\rangle$%
\end{tabular}
&
\begin{tabular}{l}
$-i\alpha |100\rangle -i\delta |011\rangle $ \\
$-i\alpha |100\rangle +i\delta |011\rangle $%
\end{tabular}
&\begin{tabular}{l}
$\sigma_x\otimes I\otimes I$ \\
$\sigma_y\otimes I\otimes I$%
\end{tabular}  \\
$|RLR\rangle$ &
\begin{tabular}{l}
$|000\rangle,|011\rangle,|101\rangle,|110\rangle$ \\
$|001\rangle,|010\rangle,|100\rangle,|111\rangle$%
\end{tabular}
&
\begin{tabular}{l}
$-i\alpha |010\rangle -i\delta |101\rangle $ \\
$-i\alpha |010\rangle +i\delta |101\rangle $%
\end{tabular}
& \begin{tabular}{l}
$I\otimes\sigma_x\otimes I$ \\
$I\otimes\sigma_y\otimes I$%
\end{tabular} \\
$|RRL\rangle$ &
\begin{tabular}{l}
$|000\rangle,|011\rangle,|101\rangle,|110\rangle$ \\
$|001\rangle,|010\rangle,|100\rangle,|111\rangle$
\end{tabular}
&
\begin{tabular}{l}
$-i\alpha |001\rangle -i\delta |110\rangle $ \\
$-i\alpha |001\rangle +i\delta |110\rangle $
\end{tabular}
& \begin{tabular}{l}
$I\otimes I\otimes \sigma_x$ \\
$I\otimes I\otimes \sigma_y$
\end{tabular}\\ \hline\hline\end{tabular}
\end{center}
\end{table}

\section{Discussions and summary}

\label{dc} Our scheme, involving only virtual excitation of the atoms, is
robust to spontaneous emission. Besides, as cavity decay has been considered
in the reflection coefficient Eq. (\ref{r}), our scheme could work with bad
cavities in the case of large cavity decay and/or weak couplings. The
dominant source of error in our scheme is the photon loss due to the cavity
mirror absorption and scattering, the fiber absorption, and the inefficiency
of the detector. Nevertheless, since the accomplishment of our scheme relies
on the successful detection of the photons, the photon loss does not affect
the fidelity, but only the efficiency.

The Hadamard gate operations on the atoms could be done using Raman
configuration by two polarized lasers detuned from the
$|0\rangle\leftrightarrow |e\rangle$ and $|1\rangle\leftrightarrow
|e\rangle$ transitions. In addition, the measurement of the atomic
states could be carried out by a resonant laser. For example, the
successful detection of a leaking photon from the cavity after the
radiation on the atom by a L-polarized laser means the population in
the atomic state $|0\rangle$. Otherwise, we may try to detect the
atomic state $|1\rangle$ using a R-polarized laser.

We argue that the implementation of our scheme heavily depends on
the imperfection rate due to relevant techniques. Under currently
available technique, our scheme could be accomplished within a
finite time if the qubit number $N$ is not very large. Specifically,
supposing that the failure rate associated with the decay of the
virtual atomic excitation is about 2$\%$, the current dark count
rate of the single-photon detector yields the inefficiency of
$10^{-4}$, and other imperfection rate due to photon loss is about
$6\%$, we thus have the success rate to be $[(1-2\%)^2 \times
10^{-4} \times (1-6\%)]^{N}$. Fortunately, thanks to the highly
efficient single-photon generator producing 10000 photons every
second \cite{photon}, we may accomplish the teleportation in finite
time. For example, a successful teleportation of a two-qubit state
takes about three hours.

\begin{figure}[t]
\begin{center}
\scalebox{0.7}{\includegraphics{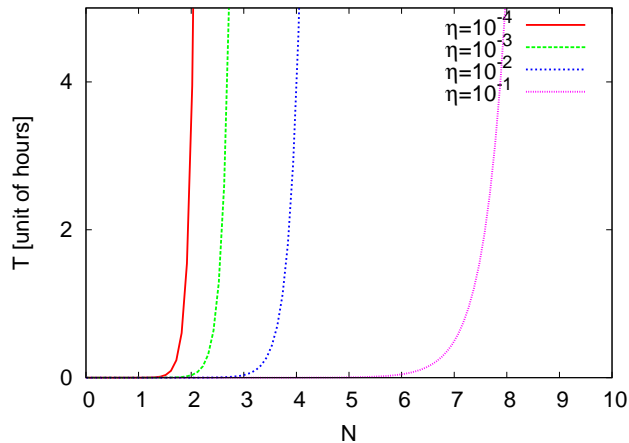}}
\end{center}
\caption{(Color online). The implementation time of the
teleportation as a function of the number of the parties involved,
with different single-photon detector inefficiency $\eta$. }
\label{time}
\end{figure}

This time-consuming implementation is mainly resulted from the low
efficiency of currently available single-photon detectors, which is
also a problem in any of the previously published schemes using
photon interference. To shorten the implementation time, on the one
hand, we should increase the resource. For example, if Bob possesses
$N$ sets of single-photon source, then the implementation time would
be reduced by $N$ times. On the other hand, if the efficiency of the
single-photon detector could be enhanced, then our implementation
time would be much reduced by several degrees of magnitude. A simple
estimate of the implementation time of our scheme is plotted in Fig.
\ref{time} with respect to different imperfect factors.

In comparison to previous proposals for teleportation considering
virtually excited cavity modes \cite{Liu04} and cavity decay
\cite{Bose99}, our scheme is advantageous to work very well not only
in the case of the bad cavities, but also in perfect and
deterministic fashion. Using photons as flying qubits to transfer
quantum information, it is more suitable for long-distance
communication compared to the schemes in Ref.
\cite{Liu04,Zheng04Cardoso05}. Moreover, our scheme using bipartite
entanglement as quantum channels is more robust to decoherence than
others based on multipartite entanglement
\cite{Rigolin05Yeo06Muralidharan08}.

In conclusions, we have proposed a practical scheme for teleportation of an
arbitrary $N$-partite pure state using the Faraday rotation. Besides the use
in atomic system, our idea could also be applied to quantum-dot system after
minor modification: Replacing the atomic excitations by the excitonic ones
\cite{Leuenberger05}. As it needs no CNOT operation, only involves
product-state measurements, and works perfectly and deterministically in
low-Q cavities, our scheme would be useful for building quantum network and
for scalable quantum computation using currently achieved microtoroidal
resonators \cite{Dayan08} or single-sided cavities \cite{Turchette95}.

\section*{Acknowledgments}

This work is supported by NNSF of China under Grant No. 10604025 and No.
10774163, as well as Gansu Provincial NSF of China under Grant No.
0803RJZA095.


\section*{References}

\begin{thebibliography}{99}
\bibitem{Bennett93} Bennett C H, Brassard G, Cr\'{e}peau C, Jozsa R, Peres A
and Wootters W K 1993 \textit{Phys. Rev. Lett.} \textbf{70} 1895.

\bibitem{Bouwmeester97} Pan J-W, Mattle K, Eibl M, Weinfurter H, Zeilinger A
and Bouwmeester D 1997 \textit{Nature} \textbf{390} 575; Boschi D, Branca S,
DeMartini F, Hardy L and Popescu S 1998 \textit{Phys. Rev. Lett.} \textbf{80}
1121; Riebe M, H\"{a}ffner H, Roos C F, H\"{a}nsel W, Benhelm J, Lancaster G
P T, K\"{o}rber T W, Becher C, Schmidt-Kaler F, James D F V and R. Blatt
2004 \textit{Nature} \textbf{429} 734; Barrett M D, Chiaverini J, Schaetz T,
Britton J, Itano W M, Jost J D, Knill E, Langer C, Leibfried D, Ozeri R and
Wineland D J 2004 \textit{Nature} \textbf{429} 737.

\bibitem{Sherson06} Sherson J F, Krauter H, Olsson R K, Julsgaard B,
Hammerer K, Cirac I and Polzik E S 2006 \textit{Nature} \textbf{443} 557.

\bibitem{Briegel98} Briegel H-J, D\"{u}r W, Cirac J I and Zoller P 1998
\textit{Phys. Rev. Lett.} \textbf{81} 5932.

\bibitem{Kimble08} Kimble H J 2008 \textit{Nature} \textbf{453} 1023.

\bibitem{Gottesman99} Gottesman D and Chuang I L 1999 \textit{Nature}
\textbf{402} 390.

\bibitem{Lee02} Lee J, Min H and Oh S D 2002 \textit{Phys. Rev.} A \textbf{66%
} 052318.

\bibitem{Fang03} Fang J, Lin Y, Zhu S and Chen X 2003 \textit{Phys. Rev.} A
\textbf{67} 014305.

\bibitem{Rigolin05Yeo06Muralidharan08} Yeo Y and Chua W K 2006 \textit{Phys.
Rev. Lett.} \textbf{96} 060502; Rigolin G 2005 \textit{Phys. Rev.} A \textbf{%
71} 032303; Muralidharan S and Panigrahi P K 2008 \textit{Phys. Rev.} A
\textbf{77} 032321.

\bibitem{Liu04} Ye L and Guo G-C 2004 \textit{Phys. Rev.} A \textbf{70}
054303.

\bibitem{Zheng04Cardoso05} Zheng S-B 2004 \textit{Phys. Rev.} A \textbf{69}
064302; Cardoso W B, Avelar A T, Baseia B and de Almeida N G 2005 \textit{%
Phys. Rev.} A \textbf{72} 045802.

\bibitem{Bose99} Bose S, Knight P L, Plenio M B and Vedral V 1999 \textit{%
Phys. Rev. Lett.} \textbf{83} 5158; Zheng S-B and Guo G-C 2006 \textit{Phys.
Rev.} A \textbf{73} 032329.

\bibitem{Dayan08} Dayan B, Parkins A S, Aoki T, Ostby E P, Vahala K J and
Kimble H J 2008 \textit{Science} \textbf{319} 1062.

\bibitem{Leuenberger05} Leuenberger M N, Flatt\'{e} M E and Awschalom D D
2005 \textit{Phys. Rev. Lett.} \textbf{94} 107401.

\bibitem{An09} An J-H, Feng M and Oh C H 2009 \textit{Phys. Rev.} A \textbf{%
79} 032303.

\bibitem{Julsgaard01} Julsgaard B, Kozhekin A and Polzik E S 2001 \textit{%
Nature} \textbf{413} 400.

\bibitem{Turchette95} Turchette Q A, Hood C J, Lange W, Mabuchi H and Kimble
H J 1995 \textit{Phys. Rev. Lett.} \textbf{75} 4710.

\bibitem{Duan04} Duan L-M and Kimble H J 2004 \textit{Phys. Rev. Lett.}
\textbf{92} 127902.

\bibitem{Walls94} Walls D F and Milburn G J 1994 \textit{Quantum Optics}
(Springer-Verlag, Berlin).

\bibitem{Hu08} Hu C Y, Young A, O'Brien J L, Munro W J and Rarity J G 2008
\textit{Phys. Rev.} B \textbf{78} 085307.

\bibitem{explain0} The initially entangled state could be prepared via
Faraday rotation by connecting two of the cavities in Alice's side \cite%
{An09}: The single photon input-output process regarding Alice's cavities
leads the state to that in Eq. (12) in Ref. \cite{An09}. Then a detection of
the $h$-polarized photon would achieve the desired state. Meanwhile, we have
to mention that, as we focus on how to teleport the entangled state, there
is no consideration in Fig. 1 (c) for the generation of the entangled state
to be teleported.

\bibitem{Chen09} Chen Q and Feng M 2009 \textit{Phys. Rev.} A \textbf{79}
064304.

\bibitem{photon} Hijlkema M, Weber B, Specht H P, Webster S C, Kuhn A and
Rempe G 2007 \textit{Nat. Phys.} \textbf{3} 253.
\end{thebibliography}
\end{document}